\documentstyle[12pt,epsf]{article}
\epsfverbosetrue
\def\PR #1 #2 #3 {Phys.~Rev.~{\bf #1}, #2 (#3)}
\def\PRL #1 #2 #3 {Phys.~Rev.~Lett.~{\bf #1}, #2 (#3)}
\def\PRD #1 #2 #3 {Phys.~Rev.~D~{\bf #1}, #2 (#3)}
\def\PLB #1 #2 #3 {Phys.~Lett.~B~{\bf #1}, #2 (#3)}
\def\NPB #1 #2 #3 {Nucl.~Phys.~{\bf B#1}, #2 (#3)}
\def\RMP #1 #2 #3 {Rev.~Mod.~Phys.~{\bf #1}, #2 (#3)}
\def\ZPC #1 #2 #3 {Z.~Phys.~C~{\bf #1}, #2 (#3)}

\newcommand{\gtap}{{\raise.3ex\hbox{$>$\kern-.75em\lower1ex\hbox{$\sim$}}}}
\newcommand{\ltap}{{\raise.3ex\hbox{$<$\kern-.75em\lower1ex\hbox{$\sim$}}}}

\begin{document}
\begin{titlepage}

\rightline{hep-ph/9705398} 
\bigskip\bigskip

\begin{center} {\Large \bf Single-Top-Quark Production via $W$-Gluon 
Fusion \\ 
\medskip at Next-to-Leading Order} \\
\bigskip\bigskip\bigskip\bigskip
{\large{\bf T.~Stelzer, Z.~Sullivan} and {\bf S.~Willenbrock}} \\ 
\medskip 
Department of Physics \\
University of Illinois \\ 1110 West Green Street \\  Urbana, IL\ \ 61801 \\
\bigskip 
\end{center} 
\bigskip\bigskip\bigskip

\begin{abstract}
Single-top-quark production via $W$-gluon fusion at hadron colliders
provides an opportunity to directly probe the charged-current
interaction of the top quark.  We calculate the next-to-leading-order
corrections to this process at the Fermilab Tevatron, the CERN Large
Hadron Collider, and DESY HERA.  Using a $b$-quark distribution
function to sum collinear logarithms, we show that there are two
independent corrections, of order $1/\ln (m_t^2/m_b^2)$ and
$\alpha_s$.  This observation is generic to processes involving a
perturbatively derived heavy-quark distribution function at an energy
scale large compared with the heavy-quark mass.
\end{abstract}

\end{titlepage}

\newpage

\section{Introduction}

Now that the existence of the top quark is firmly established \cite{TOP},
attention turns to testing its properties.  A powerful 
probe of the charged-current weak interaction of the top quark at hadron 
colliders is single-top-quark production.  The two primary processes are
quark-antiquark annihilation via a virtual $s$-channel $W$ boson 
\cite{CP,SW} and $W$-gluon fusion, which involves a virtual $t$-channel 
$W$ boson (Fig.~1) \cite{DW,Y,EP}.  Within the context of the standard model, 
these processes provide a direct measurement of the Cabbibo-Kobayashi-Maskawa
matrix element
$V_{tb}$.  Beyond the standard model, they are sensitive to new physics 
associated with the charged-current weak interaction of the top quark 
\cite{CY,CMY,ABES,DZ,LOY,S,LCHZX,DYYZ}.

\begin{figure}[tbp]
\begin{center}
\epsfxsize= 3.25in    	
\leavevmode
\epsfbox{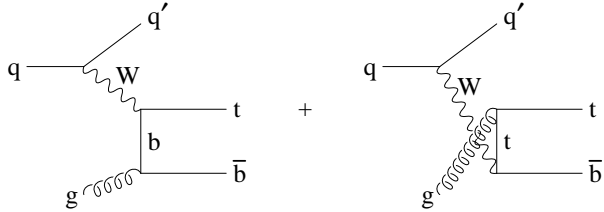}
\end{center}
\caption[fake]{Single-top-quark production via $W$-gluon fusion.}
\end{figure}

Both the precise measurement of $V_{tb}$ and the indirect detection of 
new physics require an accurate calculation of the single-top-quark production 
cross section.  The quark-antiquark-annihilation cross section has been 
calculated at next-to-leading order in QCD, with a theoretical uncertainty of
$\pm 6\%$ \cite{SmW}.  The purpose of this article is to calculate the 
next-to-leading-order correction to the $W$-gluon-fusion cross section.

A complete calculation of the next-to-leading order correction to
$W$-gluon fusion has already been presented in the literature
\cite{BV}.  However, we show that this calculation is incorrect, due
to the factorization scheme used to subtract collinear divergences.
We argue that the CTEQ $b$-quark distribution function used in that
calculation \cite{CTEQ2}, although nominally in the deep-inelastic
scattering (DIS) scheme, is actually not compatible with that scheme,
and yields incorrect results.  To avoid this problem, we perform our
calculation entirely in the modified minimal subtraction
($\overline{\rm MS}$) scheme \cite{CT}.  Our numerical results differ
significantly from those of Ref.~\cite{BV}.

We make several other contributions to the calculation 
of the next-to-leading-order correction to the $W$-gluon-fusion process:

\begin{enumerate}

\item We show that there are two independent corrections, of order
$1/\ln (m_t^2/m_b^2)$ and $\alpha_s$, which are numerically comparable.  The 
leading-order process is $qb \to q^\prime t$, as shown in Fig.~2(a). The 
$1/\ln (m_t^2/m_b^2)$ correction is associated with the diagrams in 
Figs.~2(b),~2(c), while the $\alpha_s$ correction arises from the diagrams in 
Figs.~3,4.  The existence of a correction of order $1/\ln (\mu^2/m_Q^2)$ is a 
generic feature of calculations involving perturbatively-derived 
heavy-quark distribution functions at an energy scale $\mu$ large compared with
the heavy-quark mass $m_Q$.

\item We perform the calculation in a simple and systematic way using a 
structure-function approach \cite{L,HVW}.  This allows the calculation to 
be organized in a straightfoward manner, making use of its similarity 
with deep-inelastic scattering.  

\item We carefully analyze the appropriate scale in the parton 
distribution functions.  We show that the correct scale in the 
light-quark distribution function is $\mu^2=Q^2$ ($Q^2$ is the 
virtuality of the $W$ boson), with essentially no 
scale uncertainty.  However, the appropriate scale in the $b$  
distribution function is $\mu^2 \approx Q^2 + m_t^2$.

\end{enumerate}

The paper is organized as follows.  In Sec. 2 we show that the
next-to-leading-order corrections are of two types, $1/\ln
(m_t^2/m_b^2)$ and $\alpha_s$.  We then argue that these corrections
are most reliably calculated in the $\overline{\rm MS}$ factorization
scheme.  In Sec. 3 we introduce the structure-function approach to
calculating these corrections.  In Sec. 4 we give our numerical
results and draw conclusions.  We give results for the Fermilab
Tevatron $p\bar p$ collider for $\sqrt S=$ 1.8 and 2 TeV, the CERN
Large Hadron Collider (LHC), a $pp$ collider with $\sqrt S =$ 14 TeV,
and the DESY $ep$ collider HERA with $\sqrt S=314$ GeV.  The analytic
expressions for the next-to-leading-order structure functions are
gathered in the Appendix.

\begin{figure}[tbp]
\begin{center}
\epsfxsize= 3.25in   
\leavevmode
\epsfbox{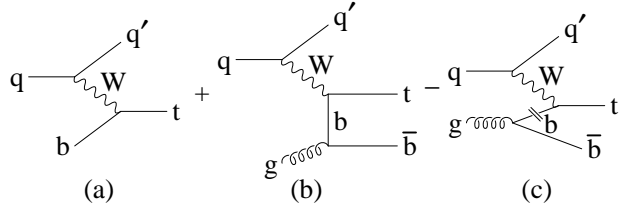}
\end{center}
\caption[fake]{(a) Leading-order process for single-top-quark production,
using a $b$ distribution function. (b) Correction to the 
leading-order process from an initial gluon. (c) Subtracting the collinear
region from (b), corresponding to a gluon splitting into a $b\bar b$ pair.
(b) and (c) taken together constitute a correction of order 
$1/\ln (m_t^2/m_b^2)$ to the leading-order process in (a).}
\end{figure}

\begin{figure}[tbp]
\begin{center}
\epsfxsize= 3.25in   
\leavevmode
\epsfbox{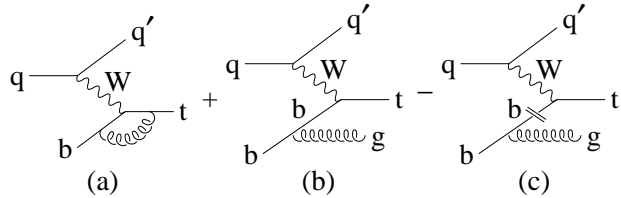}
\end{center}
\caption[fake]{Order $\alpha_s$ correction to the heavy-quark vertex in the 
leading-order
process $qb \to q^\prime t$.  (c) represents the subtraction of the 
collinear region from (b).}
\end{figure}

\begin{figure}[tbp]
\begin{center}
\epsfxsize= 3.25in    
\leavevmode
\epsfbox{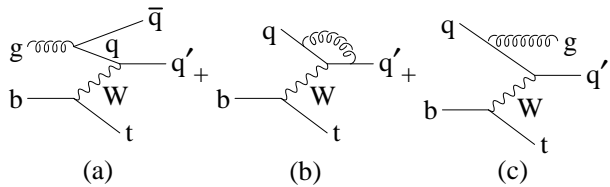}
\end{center}
\caption[fake]{Order $\alpha_s$ correction to the light-quark vertex in the 
leading-order process $qb \to q^\prime t$.}
\end{figure}
\section{Next-to-leading-order corrections}

\subsection{$1/\ln (m_t^2/m_b^2)$ correction}

The tree-level diagrams for $W$-gluon fusion are shown in Fig.~1. 
Since the $b$-quark mass is small compared with $m_t$, let us neglect it 
for the moment.  If the $b$ quark is massless, the 
first of these diagrams is singular when the final $\bar b$ quark is 
collinear with the incoming gluon.  This
kinematic configuration corresponds to the incoming gluon splitting into 
a real $b\bar b$ pair.  The propagator of the internal $b$ quark in the diagram
is therefore on-shell, and is infinite.

In reality the $b$ quark is not massless, and its mass regulates the 
collinear singularity which exists in the massless case.  The collinear 
singularity manifests itself in the total cross section as terms
proportional to $\ln [(Q^2+m_t^2)/m_b^2]$, where $Q^2 \equiv -q^2$ is the 
virtuality of the $W$ boson of four-momentum $q$.  Since the virtuality of 
the $W$ boson is controlled by the $W$ propagator, $Q^2$ is typically 
less than or of order $M_W^2$.  For readability, we write the logarithm as 
$\ln (m_t^2/m_b^2)$ in the following discussion (since $m_t^2 
\gg M_W^2$), although we use the exact expression in all calculations.

The total cross section for $W$-gluon fusion contains these
logarithmically enhanced terms, of order $\alpha_s\ln (m_t^2/m_b^2)$,
as well as terms of order $\alpha_s$ (both terms also carry a factor
of $\alpha_W^2$, which we suppress in the following discussion).
Furthermore, logarithmically enhanced terms, of order $\alpha_s^n\ln
^n(m_t^2/m_b^2)/n!$, appear at every order in the perturbative
expansion in the strong coupling, due to collinear emission of gluons
from the internal $b$-quark propagator.  Since the logarithm is large,
the perturbation series does not converge quickly, and it appears
difficult to obtain a precise prediction for the total cross section.

Fortunately, this difficulty can be obviated.  A formalism exists to
sum the collinear logarithms to all orders in perturbation theory
\cite{OT,BHS,ACOT}.  The coefficient of the logarithmically-enhanced
term is the Dokshitzer-Gribov-Lipatov-Altarelli-Parisi (DGLAP)
splitting function $P_{qg}$, which describes the splitting of a gluon
into a $b\bar b$ pair.  One can sum the logarithms by introducing a
$b$ distribution function $b(x,\mu^2)$ and calculating its evolution
with $\mu$ (from some initial condition) via the DGLAP equations.
Thus the $b$ distribution function can be regarded as a device to sum
the collinear logarithms.  Since it is calculated from the splitting
of a gluon into a collinear $b\bar b$ pair, it is intrinsically of
order $\alpha_s\ln (\mu^2/m_b^2)$.  We elaborate on this point at the
end of this section.

Once a $b$ distribution function is introduced, it changes the way one
orders perturbation theory.  The leading-order process is now $qb \to
q^\prime t$, shown in Fig.~2(a).  This cross section is of order
$\alpha_s \ln (m_t^2/m_b^2)$, due to the $b$ distribution function
($\mu \approx m_t$).  The $W$-gluon-fusion process, shown in
Fig.~2(b), contains terms of both order $\alpha_s \ln (m_t^2/m_b^2)$
and $\alpha_s$, as discussed above.  However, the logarithmically
enhanced terms have been summed into the $b$ distribution function and
thus are already present in Fig.~2(a).  It is therefore necessary to
remove these terms from the $W$-gluon fusion process to avoid double
counting.  This is indicated schematically in Fig.~2(c); the double
lines crossing the internal $b$-quark propagator indicate that it is
on-shell, which corresponds to the kinematic region responsible for
the large collinear logarithm \cite{OT,BHS,ACOT}.

After the subtraction of the terms of order $\alpha_s \ln
(m_t^2/m_b^2)$ in Fig.~2(b) by the terms in Fig.~2(c), the remaining
terms are of order $\alpha_s$.  Compared with the leading-order
process in Fig.~2(a), this is suppressed by a factor $1/\ln
(m_t^2/m_b^2)$.  Thus the diagrams of Figs.~2(b),~2(c), taken
together, correspond to a correction to the leading-order cross
section [Fig.~2(a)] of $1/\ln (m_t^2/m_b^2)$, not of order $\alpha_s$.
This is an essential point which has been previously overlooked.

This observation is generic to any process involving a perturbatively
derived heavy-quark distribution function in the region $\mu^2 \gg
m_Q^2$.  For example, the calculation analogous to the diagrams in
Figs.~2(b),~2(c) for charm production in neutral-current deep
inelastic scattering \cite{ACOT} corresponds to a correction of order
$1/\ln (Q^2/m_c^2)$ for $Q^2 \gg m_c^2$.

Let us elaborate on our contention that the $b$ distribution function
is intrinsically of order $\alpha_s \ln (\mu^2/m_b^2)$, rather than
merely of order $\alpha_s$.  If one neglects gluon bremsstrahlung and
the scale dependence of the gluon distribution function and the strong
coupling, one can solve the DGLAP equation for the $b$ distribution
function analytically [with the initial condition $b(x,\mu^2)=0$ at
$\mu=m_b$] \cite{OT,BHS,ACOT}:
\begin{equation}
b(x,\mu^2) = \frac{\alpha_s(\mu^2)}{2\pi}\ln \left(\frac{\mu^2}{m_b^2}\right) 
\int_x^1 \frac{dz}{z}
P_{qg}(z)g\left(\frac{x}{z},\mu^2\right) \;,
\label{b}
\end{equation}
where the DGLAP splitting function is given by
\begin{equation}
P_{qg}(z)=\frac{1}{2}[z^2+(1-z)^2]\;.
\end{equation}
Equation (\ref{b}) shows that $b(x,\mu^2)$ is of order $\alpha_s \ln
(\mu^2/m_b^2)$ compared with the gluon distribution function.  To
support this, we show in Fig.~5 the ratio $b(x,\mu^2)/g(x,\mu^2)\times
2\pi/\alpha_s(\mu^2)$ as a function of $\mu$ for various fixed values
of $x$, using the CTEQ4M parton distribution functions \cite{CTEQ4}.
The curves are approximately linear when $\mu$ is plotted on a
logarithmic scale, indicating that $b(x,\mu^2) \propto
[\alpha_s(\mu^2)/2\pi] \ln (\mu^2/m_b^2) g(x,\mu^2)$.

\begin{figure}[tb]
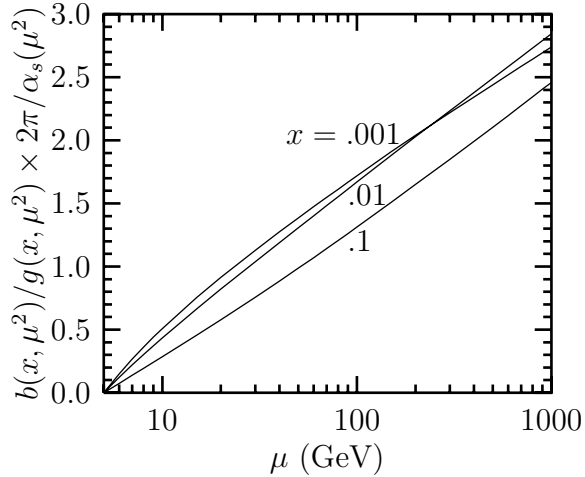

\begin{center}
\include{wgtfig5}
\end{center}
\caption[fake]{The ratio of the $b$ distribution function to the gluon 
distribution function, times $2\pi/\alpha_s(\mu^2)$, versus the factorization
scale $\mu$, for various fixed values of $x$. 
The curves are approximately linear when $\mu$ is
plotted on a logarithmic scale, indicating 
that $b(x,\mu^2) \propto [\alpha_s(\mu^2)/2\pi] \ln (\mu^2/m_b^2) g(x,\mu^2)$,
as suggested by the approximation of Eq.~(\ref{b}). }
\end{figure}

The $b$ distribution function is on a different footing from the
light-quark distribution functions.  The light-quark distribution
functions involve nonperturbative QCD, and must be measured (or
calculated nonperturbatively).  The $b$ distribution function involves
energies of order $m_b$ and larger, so it can be calculated
perturbatively; no measurement is necessary.  Given the gluon and
light-quark distributions functions, perturbative QCD makes a definite
prediction for the $b$ distribution function.

\subsection{$\alpha_s$ correction}

There are also bona fide $\alpha_s$ corrections to the leading-order
process $qb\to q^\prime t$.  The diagram in Fig.~3(a) is such a
correction; it is of order $\alpha_s^2\ln (m_t^2/m_b^2)$ [including
the factor $\alpha_s \ln (m_t^2/m_b^2)$ from the $b$ distribution
function], so it is suppressed by a factor of $\alpha_s$ with respect
to the leading-order process.

The diagram of Fig.~3(b) contains terms of both order $\alpha_s^2\ln^2
(m_t^2/m_b^2)$ and $\alpha_s^2\ln (m_t^2/m_b^2)$.  The former terms
arise from the collinear emission of the gluon, which gives rise to
another factor of $\ln (m_t^2/m_b^2)$ (on top of the factor from the
$b$ distribution function).  Similar to the discussion above, another
power of this logarithm appears at every order in the strong coupling,
and summation is required to improve the convergence of perturbation
theory.  The coefficient of this logarithmically enhanced term is the
DGLAP splitting function $P_{qq}$, which describes the splitting of a
quark into a quark and a gluon.  The collinear logarithms are summed
by adding another term, corresponding to gluon emission, to the DGLAP
evolution equation for the $b$ distribution function.  Once this is
done, the collinear region must be subtracted from Fig.~3(b); this is
shown schematically in Fig.~3(c).  The remaining terms are of order
$\alpha_s^2 \ln (m_t^2/m_b^2)$, so they are bona fide $\alpha_s$
corrections to the leading-order process.

Finally, there are the corrections to the light-quark vertex in the
leading-order process, as shown in Fig.~4.  These are also bona fide
$\alpha_s$ corrections.  Figs.~4(a),~4(b) contain collinear logarithms
$\ln (Q^2/m_q^2)$ (where $m_q$ is a light-quark mass) which are
absorbed by the light-quark distribution functions in the usual way.
Since the light-quark distribution functions are intrinsically of
zeroth order in $\alpha_s$, the remaining corrections are of order
$\alpha_s$.

\subsection{Higher orders}

Consider the next-to-next-to-leading order diagram in Fig.~6.
This diagram generates terms of
order $\alpha_s^2 \ln^2 (m_t^2/m_b^2)$, $\alpha_s^2 \ln (m_t^2/m_b^2)$,
and $\alpha_s^2$.  The term of order $\alpha_s^2 \ln^2 (m_t^2/m_b^2)$
comes from the region in which the initial gluon splits into a collinear
$b\bar b$ pair, and the $b$ quark subsequently radiates a collinear gluon.
This term is summed by the leading-order DGLAP equation, which sums
leading logarithms $\alpha_s^n \ln^n (m_t^2/m_b^2)/n!$, as discussed in 
Sec. 2.1.  Thus this term is already present in the leading-order
diagram, Fig.~2(a).

\begin{figure}[tbp]
\begin{center}
\epsfxsize= 1.5in
\leavevmode
\epsfbox{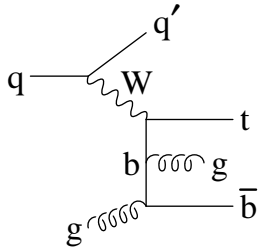}
\end{center}
\caption[fake]{Next-to-next-to-leading-order contribution to single-top-quark
production via $W$-gluon fusion.}
\end{figure}

The terms of order $\alpha_s^2 \ln (m_t^2/m_b^2)$ come from two
sources.  The first is when the initial gluon splits into a collinear
$b\bar b$ pair, and the $b$ quark subsequently radiates a noncollinear
gluon.  This is associated with the diagrams in Figs.~3(b) and 3(c),
taken together, which correspond to noncollinear gluon radiation.  The
logarithm is summed via the leading-order DGLAP equation into the $b$
distribution function in Figs.~3(b),~3(c), so this term is already
accounted for.

The other term of order $\alpha_s^2 \ln (m_t^2/m_b^2)$ is summed by
extending the DGLAP splitting function $P_{qg}$ to next-to-leading
order.  This sums the first subleading logarithms, of order
$\alpha_s^n \ln^{n-1} (m_t^2/m_b^2)$ ($n \ge 2$) into the $b$
distribution function of the leading-order process, Fig.~2(a).  The
remaining term, of order $\alpha_s^2$, is a correction of order
$\alpha_s \times 1/\ln (m_t^2/m_b^2)$ compared with the leading order
process of Fig.~2(a).

This analysis demonstrates that all collinear logarithms are
ultimately summed into the $b$ distribution function; no explicit
collinear logarithms remain.  The remaining terms are all of order
$\alpha_s^n$ or, if the diagram has a $b$ quark in the initial state,
of order $\alpha_s^n \ln (m_t^2/m_b^2)$.  These correspond to
corrections of order $\alpha_s^{n-1} \times 1/\ln (m_t^2/m_b^2)$ or
$\alpha_s^{n-1}$, respectively, compared with the leading-order
process.  For a more detailed discussion of higher orders, see
Ref.~\cite{JSMITH}.

\subsection{Factorization scheme for heavy quarks}

The factorization scheme used to eliminate the collinear divergences
from the parton cross section must be the same as the scheme used to
define the parton distribution functions in order to yield a correct
(and scheme-independent) result.  In the $\overline{\rm MS}$ scheme,
the $b$ distribution function $b(x,\mu^2)$ is defined to be zero at
$\mu=m_b$, and is then evolved to higher values of $\mu$ via the DGLAP
equations \cite{CT}.  This is the definition of the $b$ distribution
function employed in the CTEQ $\overline{\rm MS}$ parton distribution
functions \cite{CTEQ2,CTEQ4}.

Another popular factorization scheme is the DIS scheme.  In this
scheme, the neutral-current structure function $F_2(x,Q^2)$ is defined
to have no radiative correction for light quarks.  For $\mu\gg m_b$,
the $b$ quark is essentially a light quark, so a natural
interpretation of the DIS scheme for the $b$ quark is that its
contribution to $F_2(x,Q^2)$ has no radiative correction.  This is the
interpretation that was made in Ref.~\cite{BV}, which adopted the DIS
scheme for the parton cross section and used the CTEQ DIS distribution
functions \cite{CTEQ2}.  However, the CTEQ DIS $b$ distribution
function is actually not in the DIS scheme as interpreted in
Ref.~\cite{BV}.  Rather, the $b$ distribution function is again
defined by the initial condition $b(x,\mu^2)=0$ at $\mu=m_b$, and
evolved to higher values of $\mu$ via the DGLAP equations.  There is
no sense in which this yields a $b$ distribution function which is
formally equivalent to the usual DIS scheme.  As a consequence, it is
not correct to calculate the parton cross section in the usual DIS
scheme when using the CTEQ DIS $b$ distribution function.  The same is
true of the CTEQ DIS charm distribution function.

To avoid this problem, we calculate entirely in the $\overline{\rm
MS}$ scheme.  This yields very different numerical results from the
calculation of Ref.~\cite{BV} in the DIS scheme.

\section{Structure-function approach}

Inspecting the leading-order process in Fig.~2(a), $qb\to q^\prime t$,
one observes that it is analogous to charged-current deep-inelastic
scattering.  In fact, it is double deep-inelastic scattering; the
virtual $W$ boson is probing both the hadron containing the $b$ quark,
and the hadron containing the light quark, $q$.  This is shown
schematically in Fig.~7.  We can exploit this analogy to calculate the
corrections to this process in a compact way, in terms of
next-to-leading-order hadronic structure functions \cite{L,HVW}.  This
factorization of the process is exact at next-to-leading order,
because diagrams involving gluon exchange between the light-quark and
heavy-quark lines do not interfere with the tree diagram, due to color
conservation.

\begin{figure}[tbp]
\begin{center}
\epsfxsize= 2in
\leavevmode
\epsfbox{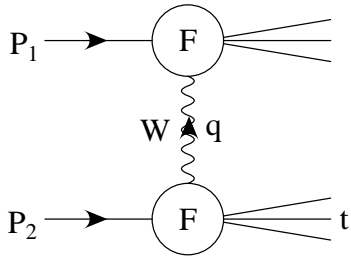}
\end{center}
\caption[fake]{Single-top-quark production via $W$-gluon fusion
from a structure-function point of view.  The $W$ boson initiates
deep inelastic scattering on both hadrons.}
\end{figure}
 
The hadronic tensor describing a $W$ boson of four-momentum
$q$ striking a hadron of four-momentum $P$ can be written in terms of 
five structure functions:
\begin{eqnarray}
MW_{\mu\nu}(x,Q^2) & = &
F_1(x,Q^2)\left(-g_{\mu\nu}+\frac{q_\mu q_\nu}{q^2}\right)
+\frac{F_2(x,Q^2)}{P\cdot q}\left(P_\mu-\frac{P\cdot q}{q^2}q_\mu \right)
\left(P_\nu-\frac{P\cdot q}{q^2}q_\nu \right) \nonumber \\
&&\mbox{} -i\frac{F_3(x,Q^2)}{2P\cdot q}\epsilon_{\mu\nu\rho\sigma}P^\rho
q^\sigma +F_4(x,Q^2)q_\mu q_\nu
+F_5(x,Q^2)(P_\mu q_\nu + P_\nu q_\mu) \;,
\end{eqnarray} 
where $Q^2 = -q^2$.  If the struck quark, and the quark into which it is 
converted, are both massless, then the current with which the $W$ interacts is
conserved, and one has $q^\mu W_{\mu\nu} = q^\nu W_{\mu\nu} = 0$.  This
implies that the structure functions $F_4,F_5$ vanish.  The scaling
variable $x$ is given by $x=Q^2/2P\cdot q$, as usual.

If the quark into which the struck quark is converted is massive, such as the 
top quark, then the current is no longer conserved, and $F_4,F_5$ are 
nonvanishing (although we will find that they do not enter our calculation).  
Furthermore, the scaling variable is now given by 
$x=(Q^2+m_t^2)/2P\cdot q$.

The hadronic cross section in Fig.~7 is obtained by contracting the hadronic
tensors at each vertex with the square of the $W$ propagator connecting them.
Due to current conservation of the light-quark tensor, the $q^\mu q^\nu/M_W^2$
term in the numerator of the $W$ propagator does not contribute, so one
simply contracts the two tensors together.  One finds
\begin{eqnarray}
\lefteqn{MW_{\mu\nu}(x_1,Q^2)MW^{\mu\nu}(x_2,Q^2) = }
\nonumber \\
& & 3F_1(x_1,Q^2)F_1(x_2,Q^2) \nonumber \\
& & \mbox{} +F_1(x_1,Q^2)F_2(x_2,Q^2)\frac{P_2\cdot
(-q)}{q^2} +F_2(x_1,Q^2)F_1(x_2,Q^2)\frac{P_1\cdot q}{q^2} \nonumber \\
& & \mbox{} +F_2(x_1,Q^2)F_2(x_2,Q^2)\frac{1}{P_1\cdot q P_2\cdot (-q)}
\left(P_1\cdot P_2 - \frac{P_1\cdot q P_2\cdot q}{q^2}\right)^2 \nonumber \\
& & \mbox{} +\frac{1}{2}F_3(x_1,Q^2)F_3(x_2,Q^2)\left(\frac{P_1\cdot P_2 q^2}
{P_1\cdot q P_2\cdot q}-1\right) \;,
\label{ww}
\end{eqnarray}
where 
\begin{eqnarray}
Q^2 & = & -q^2 \;, \\
x_1 & = & \frac{Q^2}{2P_1\cdot q} \;,\\
x_2 & = & \frac{Q^2+m_t^2}{2P_2\cdot (-q)}\;. 
\end{eqnarray}
The heavy-quark structure functions $F_4,F_5$ do not contribute to
this expression because they are the coefficients of tensors which
contain $q^\mu$, $q^\nu$, or both. These tensors give vanishing
contribution when contracted with the light-quark tensor, due to
current conservation.  The $W$ boson interacts with massless quarks in
the hadron of four-momentum $P_1$, and interacts with a $b$ quark in
the hadron of four-momentum $P_2$, as indicated in Fig.~7.  Note that
the latter hadron is probed by a $W$ boson of four-momentum $-q$,
which results in $P_2\cdot (-q)$ appearing in several places in
Eq.~(\ref{ww}).  One must also add the contribution where the $W$
boson interacts with massless quarks in the hadron of four-momentum
$P_2$ and with the $b$ quark in the hadron of four-momentum $P_1$.

The differential hadronic cross section is given by 
\cite{HVW}\footnote{This equation is obtained from Eq.~(2) of Ref.~\cite{HVW}
by setting $d\Gamma = 0$ and integrating out the four-dimensional Dirac 
$\delta$ function.}
\begin{equation}
d\sigma = \frac{1}{2S} 4 \left(\frac{g^2}{8}\right)^2 \frac{1}{(Q^2+M_W^2)^2}
MW_{\mu\nu}(x_1,Q^2)MW^{\mu\nu}(x_2,Q^2)
(2\pi)^2 \frac{1}{4S}dQ^2 dW_1^2 dW_2^2 \;,
\end{equation}
where $W_1^2 = (P_1 + q)^2$ and $W_2^2 = (P_2 - q)^2$ are the squared 
invariant masses of the 
hadron remnants (including the top quark), and $S=2P_1\cdot P_2$ is the 
square of the hadronic center-of-momentum energy.  Using
\begin{eqnarray}
2P_1\cdot q & = & W_1^2 + Q^2 \;, \\
2P_2\cdot (-q) & = & W_2^2 + Q^2 \;,
\end{eqnarray}
we can write Eq.~(\ref{ww}) in terms of the integration variables 
$Q^2, W_1^2, W_2^2$:
\begin{eqnarray}
\lefteqn{MW_{\mu\nu}(x_1,Q^2)MW^{\mu\nu}(x_2,Q^2) = }
\nonumber \\
& & 3F_1(x_1,Q^2)F_1(x_2,Q^2) \nonumber \\
& & \mbox{} - \frac{1}{2}F_1(x_1,Q^2)F_2(x_2,Q^2)\frac{W_2^2+Q^2}{Q^2}
-\frac{1}{2}F_2(x_1,Q^2)F_1(x_2,Q^2)\frac{W_1^2+Q^2}{Q^2} \nonumber \\
& & \mbox{} +F_2(x_1,Q^2)F_2(x_2,Q^2)\frac{1}{(W_1^2+Q^2)(W_2^2+Q^2)}
\left(S - \frac{(W_1^2+Q^2)(W_2^2+Q^2)}{2Q^2}\right)^2 \nonumber \\
& & \mbox{} +F_3(x_1,Q^2)F_3(x_2,Q^2)\left(\frac{SQ^2}
{(W_1^2+Q^2)(W_2^2+Q^2)}-\frac{1}{2}\right) \;,
\end{eqnarray}
where
\begin{eqnarray}
x_1 & = &  \frac{Q^2}{W_1^2 + Q^2} \;, \\
x_2 & = &  \frac{Q^2+m_t^2}{W_2^2 + Q^2} \;.
\end{eqnarray}
The physical region is given by
\begin{eqnarray}
W_1 & \ge & 0 \;, \\
W_2 & \ge & m_t \;, \\
W_1 + W_2 & \le & \sqrt S \;, \\
Q^2_{max \atop min} & = & \frac{1}{2}[S-W_1^2-W_2^2 \pm \lambda^{1/2}
(S,W_1^2,W_2^2)] \;, \\
\lambda(a,b,c) & = & a^2 + b^2 + c^2 - 2ab -2ac - 2bc \;.
\end{eqnarray}

The next-to-leading-order expressions for the structure functions are given 
in the Appendix.  We use the $\overline{\rm MS}$ scheme, for the 
reasons discussed in the previous section.  After the subtraction of the
collinear logarithms $\ln [(Q^2+m_t^2)/m_b^2]$, we set the $b$ mass to zero,
since it is small compared with the top-quark mass.\footnote{In practice, it
is simpler to set the $b$ mass to zero from the outset, and evaluate the
cross section in $N=4-2\epsilon$ dimensions.  The collinear logarithms 
appear as terms proportional to $1/\epsilon - \gamma + \ln 4\pi$, and are
subtracted in the $\overline{\rm MS}$ scheme.}
When evaluating the 
next-to-leading-order contribution to the cross section, we use the 
next-to-leading-order expression for the structure function corresponding to
the light quark or the heavy quark, but not both at the same time, as this 
would yield a contribution of next-to-next-to-leading order.

\subsection*{Factorization scale}

The similarity of the leading-order process $qb\to q^\prime t$ with deep 
inelastic scattering suggests that the relevant scale in the light-quark 
distribution function is $\mu^2=Q^2$.  If the parton distribution functions 
were extracted solely from deep-inelastic-scattering data at the same 
values of $x$ and $Q^2$ relevant to this process, this statement would be 
exactly correct, because the radiative corrections to deep-inelastic 
scattering are precisely the same as those to the light-quark vertex in
$qb\to q^\prime t$.  The latter process has 
additional radiative corrections, both to the heavy-quark vertex and 
between the two quark lines, but these are unrelated to the scale in the 
light-quark distribution function.

The actual situation is not far from the situation described above.
Most of the information on the light-quark distribution functions does
come from deep inelastic scattering, and the relevant values of $x$
and $Q^2$ are within the range of the HERA $ep$ collider: $x\sim
m_t/\sqrt S\sim 0.1$ at the Tevatron and $x\sim 0.01$ at the LHC, with
$Q^2\ltap M_W^2$.  We therefore set $\mu^2=Q^2$ in the light-quark
distribution function and refrain from varying the scale, as is
usually done to estimate the theoretical uncertainty from uncalculated
higher-order corrections.

The situation is entirely different for the scale in the $b$ distribution 
function.  The collinear logarithm that results from the diagrams in 
Figs.~2(b) and 3(b) is $\ln [(Q^2+m_t^2)/m_b^2]$.  Upon 
subtraction of the collinear region via the diagrams in Figs.~2(c) and 
3(c), the remaining logarithm is $\ln [(Q^2+m_t^2)/\mu^2]$ (see the Appendix).
The appropriate scale in the $b$ distribution function is therefore 
$\mu^2 \approx Q^2+m_t^2$.  Since the $b$ distribution is obtained from 
an entirely theoretical calculation, we vary this scale in order to 
estimate the uncertainty from uncalculated higher-order corrections.

The argument above shows that the appropriate scales in the 
light-quark and $b$-quark distribution functions are different. Although 
it may seem unfamiliar to have different scales in the parton distribution 
functions of a given hadronic process, we have shown that it is appropriate in
this case.
The appropriate scale for the production of a quark of mass $m_Q$ via
charged-current deep inelastic scattering is $\mu^2 = Q^2 + m_Q^2$, which 
yields $\mu^2 = Q^2$ for the light-quark structure function and 
$\mu^2 = Q^2 + m_t^2$ for the top-quark charged-current structure function.

\section{Results and Conclusions}

We evaluate the next-to-leading-order cross section for
single-top-quark production via $W$-gluon fusion using the latest CTEQ
$\overline{\rm MS}$ distribution functions, CTEQ4M \cite{CTEQ4}.  The
cross sections at the Tevatron (1.8 and 2 TeV) and the LHC for the sum
of $t$ and $\bar t$ production for\footnote{The current world-average
top-quark mass is $175.6\pm 5.5$ GeV \cite{TOPMASS}.}  $m_t=$ 175 GeV
are given in Table 1, assuming $V_{tb}=1$.  The leading-order cross
sections are also evaluated with the CTEQ4M distribution functions.
(When evaluated with the CTEQ4L leading-order distribution functions,
the leading-order cross sections are 1.61, 2.31, and 237~pb at the
three machines.)  The $1/\ln (m_t^2/m_b^2)$ and $\alpha_s$ corrections
are listed separately. The $1/\ln (m_t^2/m_b^2)$ correction is $-20\%$
at the Tevatron, and $-11\%$ at the LHC.  This confirms previous
calculations of this correction in the $\overline{\rm MS}$ scheme
\cite{CY,YC,HBB}. The $\alpha_s$ correction is $+12\%$ at the
Tevatron, and $+2\%$ at the LHC. The next-to-leading-order cross
section is the sum of the leading-order cross section and these two
corrections.  The fact that the $\alpha_s$ correction partially
compensates the $1/\ln (m_t^2/m_b^2)$ correction is a numerical
accident, as these are two truly independent parameters.

Also given in Table~1 is the cross section for $e^- p \to \nu_e \bar t
b$ or $e^+ p \to \bar\nu_e t \bar b$ at HERA \cite{VB,Sch,VV}.  (The
leading-order cross section is $1.21 \times 10^{-4}$~pb when evaluated
with the CTEQ4L leading-order distribution functions.)  The $1/\ln
(m_t^2/m_b^2)$ correction is $-33\%$, and the $\alpha_s$ correction is
$+36\%$.  An integrated luminosity of about 10 fb$^{-1}$ would be
needed to produce a single event.  This is unattainable given the
design luminosity of the machine (${\cal L} = 1.6 \times 10^{31}
/cm^2/s$).

\begin{table}[tbp]
\begin{tabular*}{\textwidth}{r@{}l@{\extracolsep\fill}llll} \hline \hline
\multicolumn{2}{c}{$\sqrt{S}$} & LO~(pb) & $1/\ln (m_t^2/m_b^2)$~(pb) &
$\alpha_s$~(pb) & NLO~(pb) \\ \hline
1.8& ~TeV $p\bar p$ & 1.84  & -0.39 & 0.25 & 1.70 \\
2& ~TeV $p\bar p$ &   2.67  & -0.55 & 0.32 & 2.44 \\
14& ~TeV $pp$ &       270   & -31   & 6    & 245 \\ 
314& ~GeV $ep$ & 1.02$\times 10^{-4}$ & -0.34$\times 10^{-4}$ & 
0.36$\times 10^{-4}$ & 1.04$\times 10^{-4}$ \\
\hline \hline
\end{tabular*}
\caption[fake]{Cross sections for single-top-quark production via $W$-gluon
fusion at the Tevatron, LHC, and HERA for $m_t=175$ GeV.  The cross sections
are the sum of $t$ and $\bar t$ production at the Tevatron and the LHC, and
either $t$ (positron beam) or $\bar t$ (electron beam) at HERA.   
The first column gives the leading-order
cross section [Fig.~2(a)]; the second column the correction of order
$1/\ln (m_t^2/m_b^2)$ [Figs.~2(b),~2(c)]; the third column the correction 
of order $\alpha_s$ (Figs.~3,4); and the last column the 
next-to-leading-order cross section (the sum of the first three columns).
All calculations are performed in the ${\overline {\rm MS}}$ scheme using 
CTEQ4M parton distributions functions with $\mu^2=Q^2$ for the light-quark 
vertex and $\mu^2=Q^2+m_t^2$ for the heavy-quark vertex.}
\end{table}

We argued in Sec. 2.3 that the CTEQ DIS $b$ distribution function
is incompatible with the usual DIS scheme, and yields incorrect
results.  To demonstrate this, we also perform the calculation in the
DIS scheme using CTEQ4D distribution functions.  The
next-to-leading-order cross sections at the Tevatron (1.8 and 2 TeV)
and the LHC are found to be 2.24, 3.20, 290~pb.  These differ from the
results in the $\overline{\rm MS}$ scheme by much more than the
theoretical uncertainty in that calculation, which we now estimate.

To estimate the uncertainty from uncalculated higher-order
corrections, we vary the scale in the $b$ distribution function about
the central value $\mu^2 = Q^2 + m_t^2$.  The results are shown in
Fig.~8 at the Tevatron (2 TeV) and the LHC, for both the leading-order
and next-to-leading-order cross sections, using the CTEQ4M parton
distribution functions.  The next-to-leading-order cross section is
considerably less sensitive to $\mu$, as expected. Varying $\mu$
between one-half and twice its central value yields an uncertainty in
the next-to-leading-order cross section of $\pm 5\%$ at the Tevatron
and $\pm 4\%$ at the LHC.  As discussed in Sec. 3, we do not vary
the scale in the light-quark distribution function, where $\mu^2 =
Q^2$.  Although our estimate of the theoretical uncertainty in the
cross section from uncalculated higher orders is rather small, it
would be worthwhile to pursue the calculation to the next order in
$\alpha_s$.

\begin{figure}[tb]
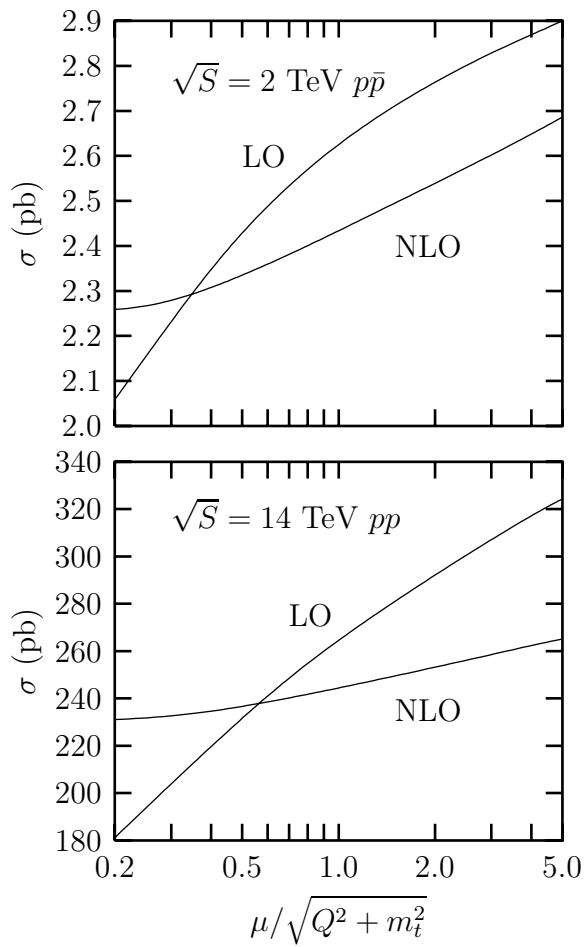

\begin{center}
\include{wgtfig8a}
\vspace*{-10ex}
\include{wgtfig8b}
\end{center}
\caption[fake]{Cross section for single-top-quark production via $W$-gluon
fusion at the Tevatron and the LHC for $m_t=175$ GeV, versus the ratio of the 
factorization scale $\mu$ to its natural value, $\mu = \sqrt {Q^2 + m_t^2}$. 
Both the leading-order and next-to-leading-order cross sections are shown.}
\end{figure}

Another source of uncertainty stems from the uncertainty in the top-quark 
mass.  The cross section as a function of the top-quark mass is shown 
in Fig.~9 at the Tevatron (2 TeV) and the LHC.  The cross section is 
relatively insensitive to the top-quark mass because the decrease in the 
parton distribution functions with increasing $m_t$ is not augmented by a
decrease in the partonic cross section, which scales like
$1/M_W^2$ instead of $1/\hat s$.  
The present uncertainty of $\pm 5.5$ GeV in the top-quark mass
\cite{TOPMASS} corresponds to an uncertainty of $\pm 9\%$ in the cross 
section at the Tevatron and $\pm 5\%$ at the LHC.  Anticipating an 
uncertainty of $\pm 2$ GeV in the top-quark mass from Run II at the Tevatron 
and/or from the LHC reduces the uncertainty in the cross section 
from the top-quark mass to $\pm 3\%$ at the Tevatron and $\pm 2\%$ 
at the LHC. 

\begin{figure}[tb]
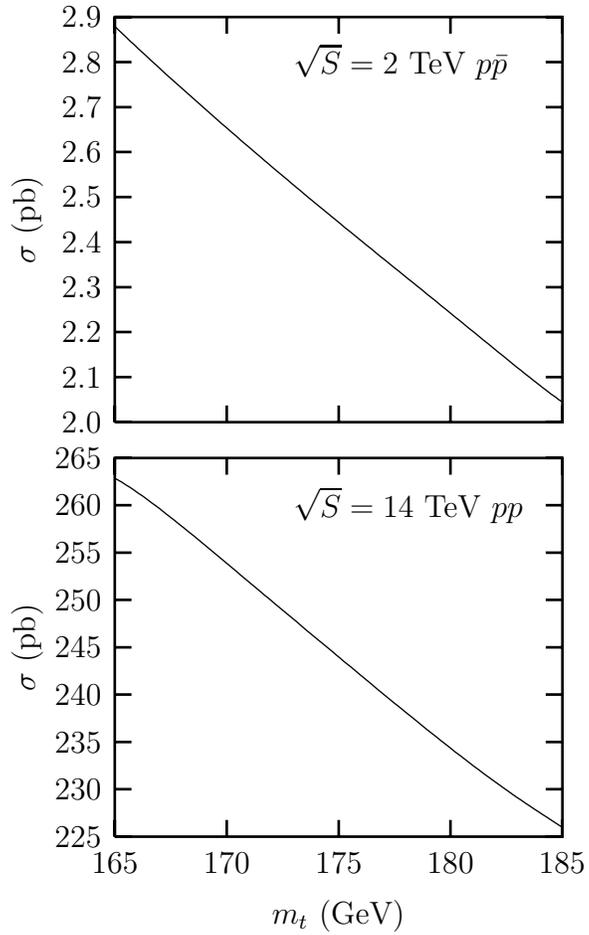

\begin{center}
\include{wgtfig9a}
\vspace*{-10ex}
\include{wgtfig9b}
\end{center}
\caption[fake]{Next-to-leading-order cross section for single-top-quark 
production via $W$-gluon 
fusion at the Tevatron and the LHC as a function of the top-quark mass.}
\end{figure}

Another source of uncertainty is the gluon distribution function, which 
reflects itself in an uncertainty in the $b$ distribution function.  It 
is impossible to estimate this uncertainty with any confidence at this 
time; what is needed is a parton distribution set with an associated
error-correlation matrix.

In this paper we present the first complete and correct calculation of the 
next-to-leading-order corrections to single-top-quark production via
$W$-gluon fusion.  We show that there are two independent corrections,
of order $1/\ln (m_t^2/m_b^2)$ and $\alpha_s$, which are numerically 
comparable.  We estimate the uncertainty due to uncalculated
higher-order corrections to be about $\pm 5\%$ at the Tevatron and the LHC. 
Assuming the uncertainty in the gluon distribution function can be quantified
and reduced to a sufficiently-small level,  
single-top-quark production via $W$-gluon fusion will be an accurate probe
of the charged-current interaction of the top quark at the Tevatron and the 
LHC.  In conjunction with $q\bar q \to t\bar b$, it will yield an accurate
measurement of $V_{tb}$ and possibly indicate the presence of new physics.

Note: The numerical results contained in this paper were obtained by 
evaluating the weak coupling constant $g$ in terms of the Fermi coupling 
$G_F$ and the $W$-boson mass $M_W$, via $g^2 = 8G_FM_W^2/\sqrt{2}$, where 
$G_F= 1.16639\times 10^{-5}$~GeV$^{-2}$ and $M_W = 80.4$~GeV.  These 
numerical results are approximately $2\%$ less than the values which appear
in the published version of this paper [Phys.~Rev.~D {\bf 56}, 5919 
(1997)].

\section*{Acknowledgements}

\indent\indent We are grateful for conversations and correspondance
with S.~Keller, F.~Olness, E.~Reya, and W.-K.~Tung. This work was
supported in part by the U.S. Department of Energy Grant
No. DE-FG02-91ER40677. We gratefully acknowledge the support of GAANN,
under Grant No. DE-P200A10532, from the U.~S.~Department of Education
for Z.~S.
 
\def\beqa {\begin{eqnarray}}
\def\eqa {\end{eqnarray}}

\newenvironment{arrayclose}{\renewcommand{\arraystretch}{1} \begin{array}}
{\renewcommand{\arraystretch}{2} \end{array}}
\renewcommand{\arraystretch}{2}

\section*{Appendix}

The structure functions for the charged-current production of a heavy
quark were calculated at next-to-leading order many years ago in
Ref.~\cite{G}.  This calculation was recently repeated in
Ref.~\cite{GKR}, which discovered a misprint in the previous result,
and also adopted the modern convention of treating the gluon as having
$N-2$ helicity states in $N$ dimensions.  We present the structure
functions below, for completeness.

Our calculation utilizes the charged-current structure functions for 
top-quark production,
$F_i(x,Q^2)$ ($i=1,2,3$), calculated in the $\overline{\rm MS}$ scheme. 
The bottom-quark mass is neglected throughout.
To make contact with Refs.~\cite{G,GKR} we define a related set of 
structure functions, ${\cal F}_i(x,Q^2)$, via $F_1 \equiv {\cal F}_1$, $F_2
\equiv 2x{\cal F}_2$, and $F_3 \equiv 2{\cal F}_3$.  These structure 
functions are related to the parton distribution functions by 
\begin{equation}
{\cal F}^q_i(x,Q^2) = q(x,\mu^2) + \frac{\alpha_s(\mu^2)}{2\pi}
\int_x^1\frac{dz}{z} \left[ H_i^q(z,Q^2,\mu^2,\lambda)\, 
q\left(\frac{x}{z},\mu^2\right) +
H_i^g(z,Q^2,\mu^2,\lambda)\, 
g\left(\frac{x}{z},\mu^2\right) \right] \;, \label{FDEFN}
\end{equation}
where 
\begin{equation}
\lambda = \frac{Q^2}{Q^2 + m_t^2}\;.
\end{equation}
The coefficient function for real and virtual gluon emission (Fig.~3) is
\beqa
H_i^q(z,Q^2,\mu^2,\lambda) &=& P_{qq}(z)\ln \frac{Q^2 + m_t^2}{\mu^2} +
h_i^q(z,\lambda) \;,
\eqa
where\footnote{The expression for $h^q$ corrects a misprint in 
Ref.~\cite{GKR}, where the $\pi^2/3$ term was written as $\pi^3/3$.}
\begin{eqnarray}
P_{qq}(z) &=& \frac{4}{3}\left( \frac{1+z^2}{1-z}\right)_+  \;, \\
&& \nonumber \\
h_i^q(z,\lambda) &=& \frac{4}{3}\left\{ h^q + A_i\delta (1-z) +
B_{1,i}\frac{1}{(1-z)_+} \right. \nonumber \\
& & \mbox{\hspace{4ex}} \left. +B_{2,i}\frac{1}{(1-\lambda z)_+} 
+ B_{3,i}\left[
\frac{1-z}{(1-\lambda z)^2}\right]_+\right\} \;, \\
&& \nonumber \\
h^q &=& - \left( 4+\frac{1}{2\lambda} +\frac{\pi^2}{3}
+\frac{1+3\lambda}{2\lambda}K_A\right) \delta(1-z) \nonumber \\
&& \mbox{} -\frac{(1+z^2)\ln z}{1-z} +(1+z^2)\left[ \frac{2\ln(1-z) -\ln(1
-\lambda z)}{1-z}\right]_+ \;, \\
&& \nonumber \\
K_A &=& \frac{1}{\lambda} (1-\lambda)\ln(1-\lambda)\;.
\end{eqnarray}
The coefficients in the expression for $h^q_i(z,\lambda)$ are given in Table 2.

\begin{table}[tbp]
\begin{tabular*}{\textwidth}{l@{\extracolsep\fill}llll} \hline \hline
$i$ & $A_i$ & $B_{1,i}$ & $B_{2,i}$ & $B_{3,i}$ \\ \hline
$1$ & $0$   & $1-4z+z^2$ & $z-z^2$  & $\frac{1}{2}$ \\
$2$ & $K_A$ & $2-2z^2-\frac{2}{z}$ & $\frac{2}{z}-1-z$ & $\frac{1}{2}$ \\
$3$ & $0$   & $-1-z^2$  & $1-z$     & $\frac{1}{2}$ \\ \hline \hline
\end{tabular*}
\caption[fake]{Coefficients in the expression for $h^q_i(z,\lambda)$.}
\end{table}

The coefficient function for initial gluons [Figs.~2(b),~2(c)] is
\beqa
H^g_{i={1,2 \atop 3}}(z,Q^2,\mu^2,\lambda) &=& P_{qg}(z)\left( \pm L_\lambda 
+\ln\frac{Q^2 + m_t^2}{\mu^2} \right) + h_i^g(z,\lambda) \;,
\eqa
where
\begin{eqnarray}
P_{qg}(z) &=& \frac{1}{2}\left[ z^2 + (1-z)^2\right] \;, \\
&& \nonumber \\
L_\lambda &=& \ln \frac{1-\lambda z}{(1-\lambda)z} \;, \\
&& \nonumber \\
h_i^g(z,\lambda) &=& C_0 + C_{1,i}z(1-z) + C_{2,i} +
(1-\lambda)zL_\lambda(C_{3,i} + \lambda z C_{4,i}) \;, \\
&& \nonumber \\
C_0 &=& P_{qg}(z)\left[ 2\ln(1-z) -\ln(1-\lambda z) -\ln z\right] \;.
\end{eqnarray}
The coefficients in the expression for $h^g_i(z,\lambda)$ are given in Table 3.

\begin{table}[tbp]
\begin{tabular*}{\textwidth}{l@{\extracolsep\fill}llll} \hline \hline
$i$ & $C_{1,i}$ & $C_{2,i}$ & $C_{3,i}$ & $C_{4,i}$ \\ \hline
$1$ & $4-4(1-\lambda)$ & $\frac{(1-\lambda)z}{1-\lambda z} -1$ & $2$ & $-4$ \\
$2$ & \begin{arrayclose}[t]{@{}l}
8-18(1-\lambda) \\
\mbox{}+12(1-\lambda)^2
\end{arrayclose} & $\frac{1-\lambda}{1-\lambda z} -1$ & $6\lambda$ &
$-12\lambda$ \\
$3$ & $2(1-\lambda)$ & $0$ & $-2(1-z)$ & $2$ \\ \hline \hline
\end{tabular*}
\caption[fake]{Coefficients in the expression for $h^g_i(z,\lambda)$.}
\end{table}

The explicit logarithms in $H^q_i(z,Q^2,\mu^2,\lambda)$ and 
$H^g_i(z,Q^2,\mu^2,\lambda)$ show that the appropriate scale for the process
is $\mu^2 = Q^2 + m_t^2$, as discussed in Sec. 3.

The structure functions for light quarks (Fig.~4)
in the $\overline{\rm MS}$ scheme
can be obtained from these expressions by taking $m_t \to 0$ $(\lambda \to 1)$.
This limit is unambiguous, except for the factor $L_\lambda$; the correct
substitution is 
 \beqa
L_{\lambda} & \rightarrow & \ln \frac{1-z}{z} \;.
\eqa


\begin{thebibliography}{99}

\bibitem{TOP} CDF Collaboration, F.~Abe~{\it et al.}, \PRL 74 2626 1995 ;
D0 Collaboration, S.~Abachi~{\it et al.}, \PRL 74 2632 1995 .

\bibitem{CP} S.~Cortese and R.~Petronzio, \PLB 253 494 1991 .

\bibitem{SW} T.~Stelzer and S.~Willenbrock, \PLB 357 125 1995 .

\bibitem{DW} S.~Willenbrock and D.~Dicus, \PRD 34 155 1986 .

\bibitem{Y} C.-P.~Yuan, \PRD 41 42 1990 .

\bibitem{EP} R.~K.~Ellis and S.~Parke, \PRD 46 3785 1992 .

\bibitem{CY} D.~Carlson and C.-P.~Yuan, \PLB 306 386 1993 .

\bibitem{CMY} D.~Carlson, E.~Malkawi, and C.-P.~Yuan, \PLB 337 145 1994 .

\bibitem{ABES} D.~Atwood, S.~Bar-Shalom, G.~Eilam, and A.~Soni, \PRD 54 
5412 1996 .

\bibitem{DZ} A.~Datta and X.~Zhang, \PRD 55 2530 1997 .

\bibitem{LOY} C.~S.~Li, R.~Oakes, and J.~M.~Yang, \PRD 55 1672 1997 ;
\PRD 55 5780 1997 .

\bibitem{S} E.~Simmons, \PRD 55 5494 1997 .

\bibitem{LCHZX} G.~Lu Y.~Cao, J.~Huang, J.~Zhang, and Z.~Xiao, hep-ph/9701406.

\bibitem{DYYZ} A.~Datta, J.~Yang, B.-L.~Young, and X.~Zhang, \PRD 56
3107 1997 .

\bibitem{SmW} M.~Smith and S.~Willenbrock, \PRD 54 6696 1996 .

\bibitem{BV} G.~Bordes and B.~van~Eijk, \NPB 435 23 1995 .

\bibitem{CTEQ2} CTEQ Collaboration, J.~Botts, J.~Morfin, J.~Owens, 
J.~Qiu, W.-K.~Tung, and H.~Weerts, \PLB 304 159 1993 .

\bibitem{CT} J.~Collins and W.-K.~Tung, \NPB 278 934 1986 .

\bibitem{L} J.~Lindfors, \PLB 167 471 1986 .

\bibitem{HVW} T.~Han, G.~Valencia, and S.~Willenbrock, \PRL 69 3274 1992 .

\bibitem{OT} F.~Olness and W.-K.~Tung, \NPB 308 813 1988 .

\bibitem{BHS} R.~Barnett, H.~Haber, and D.~Soper, \NPB 306 697 1988 .

\bibitem{ACOT} M.~Aivazis, J.~Collins, F.~Olness, and W.-K.~Tung, \PRD 50 
3102 1994 .

\bibitem{CTEQ4} CTEQ Collaboration, H.~Lai, J.~Huston, 
S.~Kuhlmann, F.~Olness, J.~Owens, D.~Soper, W.-K.~Tung, and H.~Weerts,
\PRD 55 1280 1997 .

\bibitem{JSMITH} J.~Smith, hep-ph/9708212.

\bibitem{VB} J.~van der Bij and U.~Baur, \NPB 304 451 1988 .

\bibitem{Sch} G.~Schuler, \NPB 229 21 1988 .

\bibitem{VV} J.~van der Bij and G.~J.~van Oldenborgh, \ZPC 51 477 1991 .

\bibitem{TOPMASS} R.~Raja, talk given at the {\sl XXXII Rencontres de Moriond
on Electroweak Interactions and Unified Theories},
Les Arcs, Savoie, France, March 15--22, 1997; M.~Cobal, talk given at the
{\sl Fifth Topical Seminar on the Irresistible Rise of the
Standard Model}, San Miniato, Tuscany, Italy, April 21--25, 1997.

\bibitem{YC} C.-P.~Yuan, in {\sl 6th Mexican School of Particles and Fields}, 
edited by J.~D'Olivo, M.~Moreno, and M.~Perez 
(World Scientific, Singapore, 1995), p.~16; D.~Carlson, hep-ph/9508278.

\bibitem{HBB} A.~Heinson, A.~Belyaev, and E.~Boos, \PRD 56 3114 1997 .

\bibitem{G} T.~Gottschalk, \PRD 23 56 1981 .

\bibitem{GKR} M.~Gl\"uck, S.~Kretzer, and E.~Reya, \PLB 380 171 1996 .

\end{thebibliography}
\end{document}